# Modelling Study of the Low-Temperature Oxidation of Large Methyl Esters

J. Biet, V. Warth, O. Herbinet[*], P.A. Glaude, F. Battin-Leclerc

Département de Chimie Physique des Réactions, Nancy-Université, CNRS,
ENSIC, 1 rue Grandville, BP 20451, 54001 Nancy Cedex, France

**Abstract**
This study focuses on the automatic generation by the software EXGAS of kinetic models for the oxidation of large methyl esters using a single set of kinetic parameters. The obtained models allow to well reproduce the oxidation of a n-decane / methyl palmitate mixture in a jet-stirred reactor. This paper also presents the construction and a comparison of models for methyl esters from $C_7$ up to $C_{17}$ in terms of conversion in a jet-stirred reactor and of ignition delay time in a shock tube. This comparison study showed that methyl esters larger than methyl octanoate behave similarly and have very close reactivities.

**Introduction**

Due to the depletion of fossil fuels, the development of renewable energy is more vital than ever [1,2]. The production of biofuels such as methyl esters is encouraged by the European Union in order to incorporate them in existing diesel fuels. Molecules produced from vegetable oils are methyl esters with long carbon chain as methyl palmitate ($C_{17}H_{34}O_2$). Due to the large size of the related kinetic mechanism (over 20000 reactions), it is necessary to determine the optimal model in terms of size and performance which could serve as a good surrogate model to simulate the oxidation of these large methyl esters and to investigate the difference with the combustion of alkanes induced by the ester function.

Methyl butanoate ($C_5H_{10}O_2$) has been the subject of several modelling studies. Fisher et al. [3] proposed a first detailed kinetic mechanism for the oxidation of this species in 2000. This mechanism was validated against experimental pressure data obtained in closed vessels [4]. While the experimental studies have mainly been performed under conditions corresponding to the high temperature region, this mechanism, written by analogy with n-alkanes, included all relevant pathways for both low and high temperature regions. More recently Metcalfe et al. [5] proposed a new version of the methyl butanoate mechanism based on that developed by Fisher et al. [3]. This model was validated against shock tube data. Gaïl et al. [6] proposed another version of the Fischer et al. mechanism [3] with validation using jet-stirred reactor data, variable pressure flow reactor data and opposed flow diffusion flame data. These studies led to a better understanding of the specific chemistry due to the presence of the ester group but they also demonstrated that methyl butanoate is not a good surrogate for biodiesels because the alkylic chain is too short and the influence of the ester group on the chemistry is emphasized. Herbinet et al. [7] developed a model for the oxidation of a larger methyl ester, methyl decanoate, including all pertinent reactions to low and high temperature regions. The model has been validated against limited available data. One feature of this model is its ability to reproduce the early production of carbon monoxide dioxide observed at low temperature by Dagaut et al. [8] during the experiments. The model predicts ignition delay times very close to those observed for n-decane in a shock tube and suggests that large methyl esters and alkanes of similar size have very similar reactivity. This large model was reduced and used to model extinction and ignition of laminar non premixed flames containing methyl decanoate [9].

The purpose of this paper is to present detailed kinetic mechanisms for the oxidation of several large methyl esters which have been automatically generated with the software EXGAS using a single set of kinetic parameters. The rules used for the automatic generation of these mechanisms are described in this paper and a comparison of models for methyl esters from $C_9$ up to $C_{17}$ in terms of conversion in a jet-stirred reactor and of ignition delay times at low temperature (700<T<1100K) is then performed.

**Description of the mechanisms generated with EXGAS**

The detailed kinetic mechanisms used in this study have been automatically generated by the computer package, EXGAS. This software has already been used for generating mechanisms in the case of alkanes [10-12], ethers [13] and alkenes [14]. We will recall here very shortly its main features which have already been much described and we will present the improvements of the reactions and rate constants needed to well represent the behavior of methyl esters.

*General features of EXGAS*

The system provides reaction mechanisms made of three parts:

● A comprehensive primary mechanism, where the only molecular reactants considered are the initial organic compounds and oxygen. According to the choices of the user, the reactant and the primary radicals can be systematically submitted to the different types of following elementary steps:

---

○ Unimolecular initiations involving the breaking of a C-C or a C-H bond.

○ Bimolecular initiations with oxygen to produce alkyl and •$HO_2$ radicals.

○ Oxidations of alkyl radicals with $O_2$ to form alkenes and $HO_2$• radicals.

○ Additions of alkyl (R•) and hydroperoxyalkyl (•QOOH) radicals to an oxygen molecule.

○ Isomerizations of alkyl and peroxy radicals (ROO• and •OOQOOH) involving a cyclic transition state; for •OOQOOH radicals, we consider a direct isomerization-decomposition to give ketohydroperoxides and hydroxyl radicals [15].

○ Decompositions of radicals by β-scission involving the breaking of C-C or C-O bonds for all types of radicals (for low temperature modelling, the breaking of C-H bonds is neglected).

○ Decompositions of hydroperoxyalkyl radicals to form cyclic ethers and •OH radicals.

○ Metatheses involving H-abstractions by radicals from the initial reactants.

○ Recombinations of radicals.

○ Disproportionations of peroxyalkyl radicals with $HO_2$• to produce hydroperoxides and $O_2$ (disproportionations between two peroxyalkyl radicals or between peroxyalkyl and alkyl radicals are neglected).

● A $C_0$-$C_2$ reaction base, including all the reactions involving radicals or molecules containing less than three carbon atoms [16]. The fact that no generic rule can be derived for the generation of the reactions involving very small compounds makes the use of this reaction base necessary.

● A lumped secondary mechanism, containing the reactions consuming the molecular products of the primary mechanism, which do not react in the reaction base. For reducing the number of reactants in the secondary mechanism, the molecules formed in the primary mechanism, with the same molecular formula and the same functional groups, are lumped into one unique species, without distinguishing between the different isomers [10]. The writing of the secondary reaction is made in order to promote the formation of $C_2$+ alkyl radicals, the reactions of which are already included in the primary mechanism [12].

Thermochemical data for molecules or radicals were automatically calculated and stored as 14 polynomial coefficients, according to the CHEMKIN II formalism [17]. These data were automatically calculated using software THERGAS [18], based on the group and bond additivity methods proposed by Benson [19]. In the case of esters, the bond dissociation energy of the C-H bonds carried by the carbon atom in α position from the ester function has been taken equal to 95.6 kcal.$mol^{-1}$ according to the value proposed by Luo [20] for ethyl propanoate.

The kinetic data of isomerizations, recombinations and unimolecular decompositions are calculated using software KINGAS [10] based on the thermochemical kinetics methods [19] using the transition state theory or the modified collision theory. The kinetic data, for which the calculation is not possible by KINGAS, are estimated from correlations, which are based on quantitative structure-reactivity relationships and obtained from a literature review [11].

*Generation of the primary mechanism of the oxidation of methyl esters*

The primary mechanisms have been generated using mainly the same rules as those proposed in the case of the oxidation of alkanes [11]. No new specific reaction has been considered. The changes made in the estimation of kinetic data due to the presence of ester function are the following:

● Initiations by breaking of a C–C bond.

The activation energy of unimolecular initiation which involves the breaking of the C-C bond located in the α position from the ester function has been taken equal to 93 kcal.$mol^{-1}$ according to the value recently calculated by El Nahas et al. [21]. For the other unimolecular initiations, the values proposed by El Nahas et al. [21] and those calculated using KINGAS [10] were in agreement within 2 kcal.$mol^{-1}$.

● Bimolecular initiations, oxidations and metatheses.

Bimolecular initiations and metatheses involve an H-abstraction from the initial reactant by oxygen molecules or small radicals, respectively. The oxidation of an alkyl radical deriving from a saturated ester leads to $HO_2$• radicals and to the conjugated unsaturated esters. The bond dissociation energy of the C-H bonds carried by the carbon atom just neighboring the ester function (95.6 kcal.$mol^{-1}$ [20]) is close to that of a tertiary carbon atom (e.g. 95.7 kcal.$mol^{-1}$ in isobutene [20]). The correlation used for the abstraction of an H-atom from the carbon atom located in α-position from the ester function has then been taken the same as for the abstraction of a tertiary H-atom in an alkane or an alkyl radical.

● Intramolecular isomerizations of radicals

As in our previous work [10], the activation energy is set equal to the sum of the activation energy for H-abstraction from the substrate by analogous radicals and the strain energy of the cyclic transition state. The activation energy for abstracting an H-atom from the carbon atom located in α-position from the ester function has been taken to that for the abstraction of a tertiary H-atom in an alkyl radical.

As they include seven members amongst which three oxygen atoms, the strain energy of the cyclic transition states involved in the isomerizations (1) and (2) presented in Figure 1 has been revaluated. The formation of such transition states never occurred during the oxidation of alkanes, since only cycles involving up to two oxygen atoms could be formed. The strain energy of a cyclic transition state including seven members amongst which two oxygen atoms is taken equal to 5 kcal.$mol^{-1}$ [11]. The new strain energies have been calculated by quantum methods at the CBS QB3 level of theory using Gaussian03 [22] as equal to 8.5 kcal/mol for reaction (1) and to 7.7 kcal/mol for reaction (2).

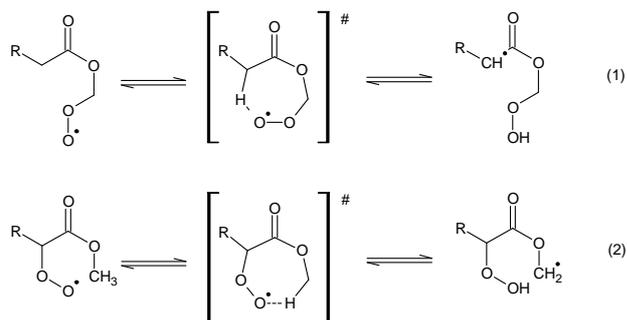

Figure 1: Example of intramolecular isomerizations involving a transition state including a seven member ring containing three oxygen atoms.

● Decompositions of radicals by β-scission

The presence of oxygen atoms in the involved radicals can have a significant impact on the values of the activation energies of the decomposition by β-scission. Quantum calculations were performed for model reactions at the CBS QB3 level of theory using Gaussian03 [22] in order to estimate these data with a given accuracy of ±1 kcal.mol$^{-1}$. Intrinsic Reaction Coordinate (IRC) calculations have been systematically performed at the B3LYP/6 31G(d) level of theory on transition states, to ensure that they are correctly connected to the desired reactants and products. Table II displays the values which have been used for the activation energies of the decompositions of oxygenated radicals by β scission.

Table 2: Activation energies (kcal.mol$^{-1}$) used for the decompositions by β-scission of oxygenated radicals involved in the oxidation of methyl esters

| Types of reaction | $E_a$ |
|---|---|
| R/C(//O)/O$^•$ → R$^•$ + CO$_2$ $^a$ | 5.1 |
| R/C(//O)/O/CH$_2$$^•$ → R/C$^•$//O + CH$_2$//O $^a$ | 31.9 |
| R/CH$^•$/C(//O)/O/CH$_3$ → R/CH//C//O + CH$_3$O$^•$ $^b$ | 49.0 |
| R/CH$^•$/CH$_2$/C(//O)/O/CH$_3$ → R/CH//CH$_2$ + CH$_3$O/C$^•$//O $^b$ | 30.7 |
| R/CH$_2$/C$^•$//O → R$^•$ + CH$_2$//C//O $^c$ | 39.9 |
| R/CH$^•$/CH$_2$/C(//O)/O/CH$_3$ → H$^•$ + R/CH//CH/C(//O)/O/CH$_3$ $^b$ | 34.9 |

$^a$ The calculation has been performed for R$^•$ = CH$_3$$^•$,
$^b$ The calculation has been performed for R$^•$ = H$^•$,
$^c$ The calculation has been performed for R$^•$ = C$_2$H$_5$$^•$.

*Generation of the secondary mechanism of the oxidation of methyl esters*

The primary mechanism of the oxidation of esters leads to the formation of hydroperoxides, cyclic ethers, ketones and aldehydes bearing an ester function, as well as of unsaturated methyl esters. New rules have been implemented in EXGAS so that reactions consuming these species are generated. These rules are described in Tables 3 to 6. The general idea followed to write these rules is the same as that used for the secondary species obtained during the oxidation of alkanes [12]. It is to promote the formation of radicals, the reactions of which are already included in the primary mechanism, i.e. C$_2$+ alkyl radicals, but also C$_2$+ methyl ester alkyl radicals.

Starting from a decomposition by breaking of a O-OH bond (Table 3), the species produced are carbon monoxide, carbon dioxide, formaldehydes, •OH and alkyl radicals from ketohydroperoxide methyl esters, aldehydes, •OH and methyl ester alkyl radicals from hydroperoxides methyl esters, acrolein and •OH and methyl ester alkyl radicals from unsaturated hydroperoxide methyl esters (>C$_4$). While alkyl or methyl ester alkyl radicals of different sizes can be obtained from the decomposition of hydroperoxides, we consider here only the formation of those containing about half of the atoms of carbon present in the molecules for which the secondary reaction is written.

Table 3: Rules for the generation by EXGAS of the reactions of acyclic hydroperoxides formed during the oxidation of methyl esters. Rate constants are given in cm$^3$, s$^{-1}$, mol units.

| Value of n | Products | Rate Parameters |
|---|---|---|
| *Decomposition of hydroperoxide methyl esters (C$_n$H$_{2n}$O$_4$PS)* | | |
| n is even (n ≥ 4) | •OH + •C$_{(n/2)}$H$_{(n-1)}$O$_2$S + C$_{(n-2)/2}$H$_{(n-1)}$CHO | 1.5.10$^{16}$ exp (-21640/T) |
| n is odd (n ≥ 5) | •OH + •C$_{(n+1)/2}$H$_n$O$_2$S + C$_{(n-3)/2}$H$_{(n-2)}$CHO | 1.5.10$^{16}$ exp (-21640/T) |
| *Decomposition of hydroperoxide unsaturated methyl esters (C$_n$H$_{(2n-2)}$O$_4$PZS)* | | |
| n = 4 | HCHO + CO$_2$ + •OH + •C$_2$H$_3$ | 1.5.10$^{16}$ exp (-21640/T) |
| n ≥ 5 | C$_2$H$_3$CHO + •OH + •C$_{(n-3)}$H$_{(2n-7)}$O$_2$S | 1.5.10$^{16}$ exp (-21640/T) |
| *Decomposition of ketohydroperoxide methyl esters (C$_n$H$_{(2n-2)}$O$_5$KPS)* | | |
| n ≥ 4 | HCHO + CO + CO$_2$ + •OH+ •C$_{(n-3)}$H$_{(2n-5)}$ | 1.5.10$^{16}$ exp (-21640/T) |

Nomenclature : S : ester (C-CO-O-C) ; P : hydroperoxide (-O-O-H) ; U : -O-O ; E : ether (C-O-C) ; E# : cyclic ether ; A : aldehyde (-CHO) ; K : ketone (C-CO-C) ; Z : double bond (C=C).

As distinguished in Table 4, two types of ethers can be obtained during the oxidation of methyl esters, ethers with the ester function inside the ring (ES#x) and ethers with the ester function outside the ring (E#xS). Both types of ethers can react by H-abstraction and then by decomposition of the obtained radicals to give oxygenated radicals and, carbon dioxide and an alkenes in the first case, and an unsaturated ester for the second one. The cyclic ether radicals obtained by H-abstractions from cyclic ethers with the ester function outside the ring and including the same number of carbon atoms as the reactant and more than 3 atoms in the cycle can still react with oxygen molecules and ultimately lead to cyclic ether hydroperoxides or be decomposed. The decomposition of cyclic ether hydroperoxides with an ester function outside the ring (> C5) leads to unsaturated esters, carbon dioxide, •OH and •HCO radicals.

Aldehydes and ketones bearing an ester function (Table 5) react by H-abstractions and lead to methyl ester alkyl radicals and, to carbon monoxide in the first case, and to ketene molecules in the second case.

Table 4: Rules for the generation by EXGAS of the reactions of cyclic ethers formed during the oxidation of methyl esters and of derived species. Rate constants are given in $cm^3$, $s^{-1}$, mol units. Small radicals involved in H-abstractions are •H, •OH, •HO$_2$, •CH$_3$, •CH$_3$OO, •C$_2$H$_5$.

| Value of n | Obtained products | Rate Parameters |
|---|---|---|
| *H-abstractions from cyclic ethers with the ester function inside the ring ($C_nH_{(2n-2)}O_3ES\#x$)[1]* | | |
| n ≥ 4 | •$C_nH_{(2n-3)}O_3ES\#x$ | As for the abstraction 6 secondary H-atoms from an alkane |
| *H-abstractions from cyclic ethers with the ester function inside the ring ($C_nH_{(2n-2)}O_3E\#xS$)* | | |
| n ≥ 5 | •$C_nH_{(2n-3)}O_3E\#xS$ | As for the abstraction 6 secondary H-atoms from an alkane |
| *Decomposition of cyclo-ether-ester radicals (•$C_nH_{(2n-3)}O_3ES\#x$)* | | |
| n ≥ 4 | $CO_2$ + •CHO + $C_{(n-2)}H_{(2n-4)}$ | $5.0.10^{13}$ exp (-12480/T) |
| *Decomposition of ester cyclo-ether radicals (•$C_nH_{(2n-3)}O_3E\#xS$)* | | |
| n = 5 | •CHO + $C_4H_6O_2S$ | $5.0.10^{13}$ exp (-12480/T) |
| n ≥ 6 | •$CH_2CHO$ + $C_{(n-2)}H_{(2n-6)}O_2S$ | $5.0.10^{13}$ exp (-12480/T) |
| *Addition of $O_2$ on ester cyclo-ether radicals (•$C_nH_{(2n-3)}O_3E\#xS$), only if x > 3* | | |
| n ≥ 5 | •$C_nH_{(2n-3)}O_5E\#xUS$ | $3.0.10^{19}$ $T^{-2.5}$ |
| *Izomerisation of peroxy ester cyclo-ether radicals (•$C_nH_{(2n-3)}O_5E\#xUS$)* | | |
| n ≥ 5 | •$C_nH_{(2n-3)}O_5E\#xPS$ | $5.0.10^{22}$ $T^{-2.5}$ exp(-20130/T) |
| *Addition of $O_2$ on hydroperoxide ester cyclo-ether radicals (•$C_nH_{(2n-3)}O_5E\#xPS$), only if x > 3* | | |
| n ≥ 5 | •$C_nH_{(2n-3)}O_7E\#xUPS$ | $8.0.10^{13}$ exp (-12830/T) |
| *Decomposition of peroxy hydroperoxide ester cyclo-ether radicals (•$C_nH_{(2n-3)}O_7E\#xUPS$)* | | |
| n ≥ 5 | •OH + $C_nH_{(2n-4)}O_6E\#xKPS$ | $1.0.10^9$ exp(-3770/T) |
| *Decomposition of ketohydroperoxide ester cyclo-ether ($C_nH_{(2n-4)}O_6E\#xKPS$)* | | |
| n = 5 | •OH + •$CH_3$ + $CH_2CO$ + 2 $CO_2$ | $5.0.10^{13}$ exp (-12480/T) |
| n ≥ 6 | •OH + $CO_2$ + •CHO + $C_{(n-2)}H_{(2n-6)}O_2S$ | $5.0.10^{13}$ exp (-12480/T) |

[1] x is the size of the ring.

The number of possible secondary reactions is larger in the case of unsaturated esters (Table 6): they can react by H abstractions or by additions of small radicals. The formation of oxygenated species and of 1,3-butadiene molecules or vinyl or allyl radicals are obtained by H-abstractions followed by decomposition.

The addition of H atoms leads to the formation of methyl ester alkyl radicals. The addition of •OH radicals produces formaldehydes and methyl ester alkyl radicals. Propene and methyl ester alkyl radicals are formed by addition of •$CH_3$ radicals.

Table 5: New rules for the generation by EXGAS of the reactions of ketones, aldehydes and alcohols formed during the oxidation of methyl esters. Rate constants are given in $cm^3$, $s^{-1}$, mol units. Small radicals involved in H-abstractions are •H, •OH, •HO$_2$, •CH$_3$, •CH$_3$OO, •C$_2$H$_5$.

| Value of n | Obtained products | Rate Parameters |
|---|---|---|
| *H-abstraction by small radicals for ester with aldehyde function ($C_nH_{(2n-2)}O_3AS$)* | | |
| n ≥ 4 | CO + •$C_{(n-1)}H_{(2n-3)}O_2S$ | As for the abstraction of 3 primary alkylic H-atoms from initial alkane |
| | CO + •$C_{(n-1)}H_{(2n-3)}O_2S$ | As for the abstraction of (2n-6) secondary alkylic H-atoms from initial alkane |
| | CO + •$C_{(n-1)}H_{(2n-3)}O_2S$ | As for the abstraction of 1 **aldehydic** H-atoms from initial alkane |
| *H-abstraction by small radicals for ester with ketone function ($C_nH_{(2n-2)}O_3KS$)* | | |
| n ≥ 4 | $CH_2CO$ + •$C_{(n-2)}H_{(2n-5)}O_2S$ | As for the abstraction of 6 primary alkylic H-atoms from initial alkane |
| | $CH_2CO$ + •$C_{(n-2)}H_{(2n-5)}O_2S$ | As for the abstraction of (2n-8) H secondary alkylic H-atoms from initial alkane |

Table 6: New rules for the generation by EXGAS of the reactions of unsaturated methyl esters. Rate constants are given in $cm^3$, $s^{-1}$, mol units. Small radicals involved in H-abstractions are •H, •OH, •HO$_2$, •CH$_3$, •CH$_3$OO, •C$_2$H$_5$.

| Value of n | Obtained products | Rate Parameters |
|---|---|---|
| *H-abstraction by small radicals for unsaturated methyl esters ($C_nH_{(2n-2)}O_2ZS$)* | | |
| n = 4 | HCHO + CO + •$C_2H_3$ | As for the abstraction of 3 primary alkylic H-atoms from initial alkane |
| n = 5 | HCHO + CO + •$C_3H_5$ | As for the abstraction of 3 primary alkylic and of 2 secondary allylic H-atoms from initial alkane |
| | HCHO + CO + $C_4H_6$ + •$C_{(n-6)}H_{(2n-11)}$ | As for the abstraction of 3 primary alkylic H-atoms from initial alkane |
| n ≥ 6 | $C_4H_6$ + •$C_{(n-4)}H_{(2n-9)}O_2S$ | As for the abstraction of 2 H secondary allylic H-atoms from initial alkane |
| | •$C_2H_3$ + $C_{(n-2)}H_{(2n-6)}O_2S$ | As for the abstraction of (2n –10) secondary alkylic H-atoms from initial alkane |
| *Additions on unsaturated methyl esters ($C_nH_{(2n-2)}O_2ZS$)* | | |
| •H | •$CH_2$-$C_{(n-1)}H_{(2n-3)}O_2S$ | $1.32.10^{13}$ exp(-1640/T) |
| | $CH_3$-•CH-$C_{(n-2)}H_{(2n-5)}O_2S$ | $1.32.10^{13}$ exp(-785/T) |
| •OH | HCHO + •$C_{(n-1)}H_{(2n-3)}O_2S$ | $2.74.10^{12}$ exp(520/T) |
| •$CH_3$ | $C_3H_6$ + •$C_{(n-2)}H_{(2n-5)}O_2S$ | $1.69.10^{11}$ exp(-3720/T) |
| | | $9.64.10^{10}$ exp(-4030/T) |

**Comparison of the reactivity of large esters**

The software EXGAS was used to generate models for the oxidation of methyl esters from $C_9$ to $C_{17}$ by using the rules described above. Because of the ester group which breaks the molecule symmetry, models for the oxidation of methyl esters are much larger than models for the oxidation of n-alkanes. As an example the model generated for methyl octanoate ($C_9H_{18}O_2$) is made of 668 species and 3905 reactions against 423 species and 2709 reactions for n-nonane ($C_9H_{20}$). As shown in Table 7 numbers of species and reactions increase drastically with the size of the molecule making simulations very time consuming for the largest methyl esters. These models are so large that they prevent their use for the CFD data tabulation.

Table 7: Numbers of species and reactions in the models generated with EXGAS.

|  | $C_9H_{18}O_2$ | $C_{11}H_{22}O_2$ | $C_{13}H_{26}O_2$ | $C_{15}H_{30}O_2$ | $C_{17}H_{34}O_2$ |
| --- | --- | --- | --- | --- | --- |
| species | 668 | 1159 | 1880 | 2877 | 4222 |
| reactions | 3905 | 6956 | 11790 | 28791 | 41573 |

Very little experimental data are available for the validation of models in the case of large esters. Experimental conversion of methyl palmitate (blended with n-decane) in a jet-stirred reactor was compared with the computed conversions for a blend of n-decane and methyl dodecanoate (Figure 2). No attempt was made to run simulations with the model for the n-decane/methyl palmitate blend because they would have been very time consuming. Experiments were carried out at a pressure of 106 kPa, a residence time of 1.5 s and at stoichiometric conditions. The mole composition of the blend was 0.74 n-decane and 0.26 methyl palmitate. The initial mole fraction of the fuel was 0.002. For the simulations the initial mole fraction of n-decane was the same than in the experiments and the initial mole fraction of methyl dodecanoate was adjusted to have the same carbon content. The agreement between experiment and modelling is satisfactorily for both fuels.

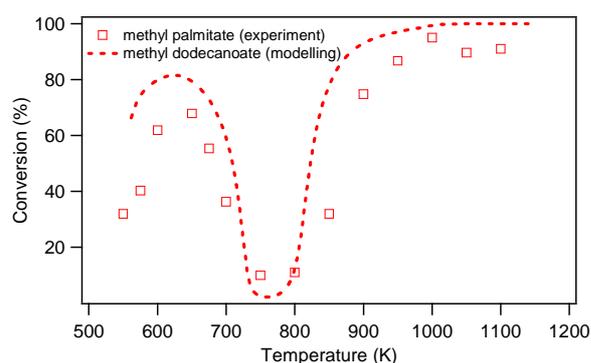

Figure 2: Comparison of experimental conversions of methyl palmitate in a jet stirred reactor with computed conversions with methyl dodecanoate.

Models generated with EXGAS were used to perform simulations in order to highlight the influence of the size of the alkyl chain on the reactivity of methyl esters. Two sets of computation are proposed here:

1) comparison of ignition delay times over a range of temperature including the negative temperature coefficient (NTC) region,
2) comparison of conversions of methyl esters in a jet-stirred reactor.

*Comparison of ignition delay times of methyl esters from $C_9$ to $C_{17}$*

Ignition delay times for methyl esters from $C_9$ to $C_{17}$ have been computed with the models generated with EXGAS (Figure 3). Simulations were carried out at temperatures from 700 to 1000 K, 13.5 bar initial pressure, at stoichiometric conditions in air. Initial mole fraction of methyl decanoate was set to 0.0134 and mole fractions of larger fuels were calculated so that the carbon content of each reacting mixture was identical.

It can be seen in Figure 3 that all methyl esters have very similar behaviors and that all models predict a S shape due to the NTC in the temperature range 750 – 1000 K. As for n-alkanes [12], the computed ignition delay times of all the studied esters are very close for temperatures below 750 K and above 1000 K. Some differences are visible in the NTC region. First the ignition delay times decrease when the size of the alkyl chain goes up (at 800 K the ignition delay time for methyl hexadecanoate is $2.7.10^{-3}$ s against $5.5.10^{-3}$ s for methyl octanoate). The second main difference is the position of the end of the NTC region which is shifted towards higher temperature when the size of the ester goes up.

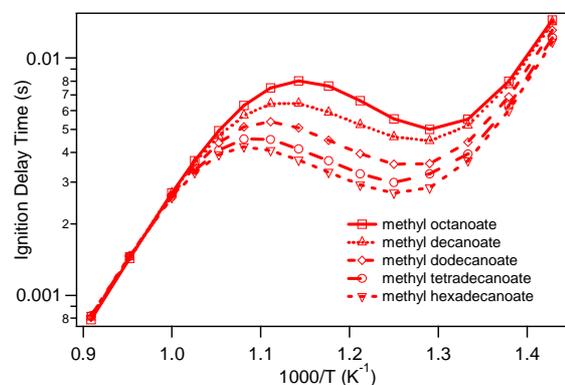

Figure 3: Computed ignition delay times for methyl esters from $C_9$ to $C_{17}$.

*Comparison of conversions of methyl esters in a jet-stirred reactor*

Conversions of neat methyl esters in a jet-stirred reactor were computed at temperatures ranging from 560 to 1200 K, at a pressure of 106 kPa and at stoichiometric conditions diluted in helium. In the case of methyl decanoate the inlet mole fraction of the fuel was 0.0023. Inlet mole fractions of other methyl esters were calculated in order to keep constant the carbon content of each reacting mixture. It can be seen in Figure 4 that all these methyl esters exhibit a very similar behaviour with the S shape due to the NTC.

Reactivities are slightly different at low temperature (below 750 K) and very close at high temperature (above 750 K). Methyl esters larger than methyl decanoate seem to be good surrogates for esters in biodiesel fuels.

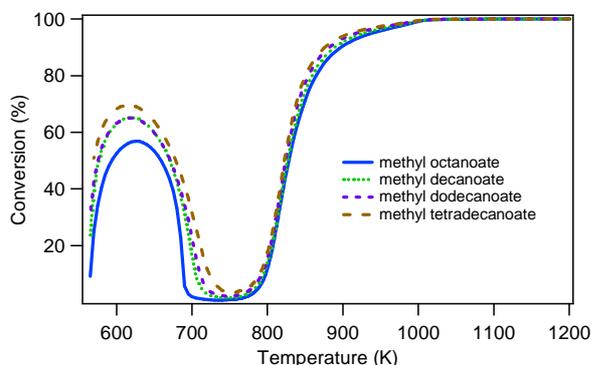

Figure 4: Computed conversions for methyl esters from $C_9$ to $C_{15}$ (jet-stirred reactor).

**Conclusions**

Detailed kinetic mechanisms for the oxidation of several large methyl esters were automatically generated with the software EXGAS using a single set of kinetic parameters. Models have been generated for methyl esters from $C_9$ to $C_{17}$. These models contain many species and reactions because of the size of the reactants and because of the loss of the symmetry due to the ester function.

The model generated for methyl octanoate was compared with available jet-stirred reactor data. The agreement between computed and experimental results was satisfactorily for the conversion of the reactant and the mole fraction of most reaction products. Models generated for methyl esters from $C_9$ to $C_{17}$ were compared in order to highlight the influence of the size of the alkyl chain on the reactivity. This comparative study showed that methyl esters larger than methyl octanoate behave very similarly with some slight differences in the NTC region. As a result, species such as methyl decanoate or methyl dodecanoate could be considered as correct surrogates for saturated methyl esters in real biodiesel fuels.

**Acknowledgment**

This work has been supported by the ADEME in collaboration with Institut Français du Pétrole, PSA Peugeot Citroën and TOTAL.